# Measuring the bending rigidity of microbial glucolipid (biosurfactant) bioamphiphile self-assembled structures by neutron spin-echo (NSE): interdigitated vesicles, lamellae and fibers

Niki Baccile,[a,*] Vincent Chaleix,[b] Ingo Hoffmann[c]

[a] Sorbonne Université, Centre National de la Recherche Scientifique, Laboratoire de Chimie de la Matière Condensée de Paris, LCMCP, F-75005 Paris, France

[b] Université de Limoges, Faculté des sciences et techniques, Laboratoire LABCiS - UR 22722, 87060 Limoges

[c] Institut Laue-Langevin, 38042 Grenoble, France

**\* Corresponding author:**
Dr. Niki Baccile
E-mail address: niki.baccile@sorbonne-universite.fr
Phone: +33 1 44 27 56 77

**Abstract**

Bending rigidity, *k*, is classically measured for lipid membranes to characterize their nanoscale mechanical properties as a function of composition. Widely employed as a comparative tool, it helps understanding the relationship between the lipid's molecular structure and the elastic properties of its corresponding bilayer. Widely measured for phospholipid membranes in the shape of giant unilamellar vesicles (GUVs), bending rigidity is determined here for three self-assembled structures formed by a new biobased glucolipid bioamphiphile, rather associated to the family of glycolipid biosurfactants than phospholipids. In its oleyl form, glucolipid G-C18:1 can assemble into vesicles or crystalline fibers, while in its stearyl form, glucolipid G-C18:0 can assemble into lamellar gels. Neutron spin-echo (NSE) is employed in the *q*-range between 0.3 nm$^{-1}$ (21 nm) and 1.5 nm$^{-1}$ (4.1 nm) with a spin-echo time in the range of up to 500 ns to characterize the bending rigidity of three different structures (Vesicle suspension, Lamellar gel, Fiber gel) solely composed of a single glucolipid. The low (*k*= 0.30 ± 0.04 $k_bT$) values found for the Vesicle suspension and high values found for the Lamellar (*k*= 130 ± 40 $k_bT$) and Fiber gels (*k*= 900 ± 500 $k_bT$) are unusual when compared to most phospholipid membranes. By attempting to quantify for the first time the bending rigidity of self-assembled bioamphiphiles, this work not only contributes to the fundamental understanding of these new molecular systems, but it also opens new perspectives in their integration in the field of soft materials.

**Introduction**

In the field of biological membranes, bending rigidity, *k*, is an intrinsic parameter that quantifies the energy required to change the curvature of membrane, and in particular the energy cost required to compress and expand the inner and outer leaflets upon bending the bilayer. [1–3] This physical property is intrinsic of the membrane and generally accepted as being dependent on its composition.[2,4] In physical biology, lipid composition determines the elastic properties of the cells, [5,6] which in turn influence their biological functions and even their preferred living environment. [7,8] Similarly, the elasticity of synthetic biomembranes in giant and small unilamellar vesicles is strongly influenced by their corresponding lipid formulation, as well as by external physicochemical conditions like ionic strength, [9] pH [10] or multilamellar structure, [3,11] but also the length scale at which *k* is measured. [6] The interplay of these parameters eventually determines the membrane stability and fields of application.[12] Studying the bending rigidity of biomembranes then constitutes a critical step in their characterization for future applications.

We address here the evaluation of the bending rigidity of a set of self-assembled membranes solely composed of a new class of amphiphiles. Known in the literature as biosurfactants,[13] the glycolipids studied in this work are produced by a fermentative process employing vegetable oils and glucose.[14,15] Even if they are referred to as surfactants, many of these molecules actually assembled into a wider variety of structures, like flat membranes, vesicles or crystalline fibers.[16] Their properties should then also be compared to phospholipids or low-molecular weight amphiphile gelators. In this regard, we measure in this work, for the first time in the field of microbial bioamphiphile biosurfactants, the bending rigidity of three non-micellar structures, flat lamellae, vesicles and crystalline fibers, entirely composed of a monounsaturated (C18:1) (vesicles and fibers) and saturated (C18:0) (lamellae) glucolipids obtained by a fermentative process.

In previous works,[17–19] it was shown that G-C18:1 has a rich and complex phase behavior:[20] G-C18:1 forms stable vesicle suspension below pH 6.2 and with a diameter in the order of several hundred nanometers (*2* in Figure 1)[17,18] as well as fibrous hydrogels (*1* in Figure 1)[19,21] when its micellar phase is put in contact with a source of calcium ions above pH 7. On the other hand, the G-C18:0 lamellar system[17,18] forms hydrogels with a defectuous structure (*3* in Figure 1),[22] the micro-macroscale behavior of which[22,23] being very much different than G-C18:1 hydrogels.[19,24]

Interestingly, both G-C18:1 and G-C18:0 are bolaform amphiphiles that can assemble into interdigitated membranes.[17,18] These, also known in the literature as monolayer lipid membranes, are known to be composed of bolaform lipids[25] in archaea and other extremophile microorganisms.[25–27] Interestingly, the bolaform structure of the lipids,[27–29] combined with their structural variability,[26] could help understanding the impressive viability of microorganisms under extreme conditions of temperature (~ -10°/+100°C), pH or pressure.[26] In this respect, even if glycolipids bioamphiphiles, such as G-C18:1 or G-C18:0, are not reported in extremophiles, they are still a product of fermentation.[30] Studying the broad panel of their physicochemical properties could then help better understanding their role in biology. From a broader *soft matter* point of view, studying the dynamics of both glycolipids assembled in three different structures, characterized by different bulk properties, is crucial to better understand this family of compounds, to put them in perspective to more classical amphiphiles and, eventually, to better identify their application potential.

Bending rigidity is classically measured with a wide panel of approaches using optical microscopy (fluctuation spectroscopy), mechanical deformation (micropipette aspiration or electrodeformation) and scattering (including small angle X-ray scattering from multibilayer stacks and neutron spin-echo).[2,31] Nonetheless, mechanical deformation methods are certainly the most common, their drawback being the need of giant unilamellar vesicles, due to the use of optical microscopy is the analytical process. In the present case, systems 1 and 3 in Figure 1 are not vesicular and the size of vesicles in system 2 is below the resolution of optical microscopes. For this reason, we employ here neutron spin-echo (NSE)[32] as a preferential technique to measure bending rigidity of lamellae, vesicles and fibers composed of single-glucose fatty acid bioamphiphiles. Despite its complexity and limited access, NSE has been long used[33,34] to determine the bending rigidity of both biological[6] and synthetic membranes, including the effects of composition,[31] charge density,[35] segregation[36] or number of lamellar stacks.[11]

In this work, NSE is preferred over mechanical deformation methods because of its major advantage of being adaptable to study bulk materials prepared as colloidal suspensions or gels and with no limitations on the minimum size and morphology.

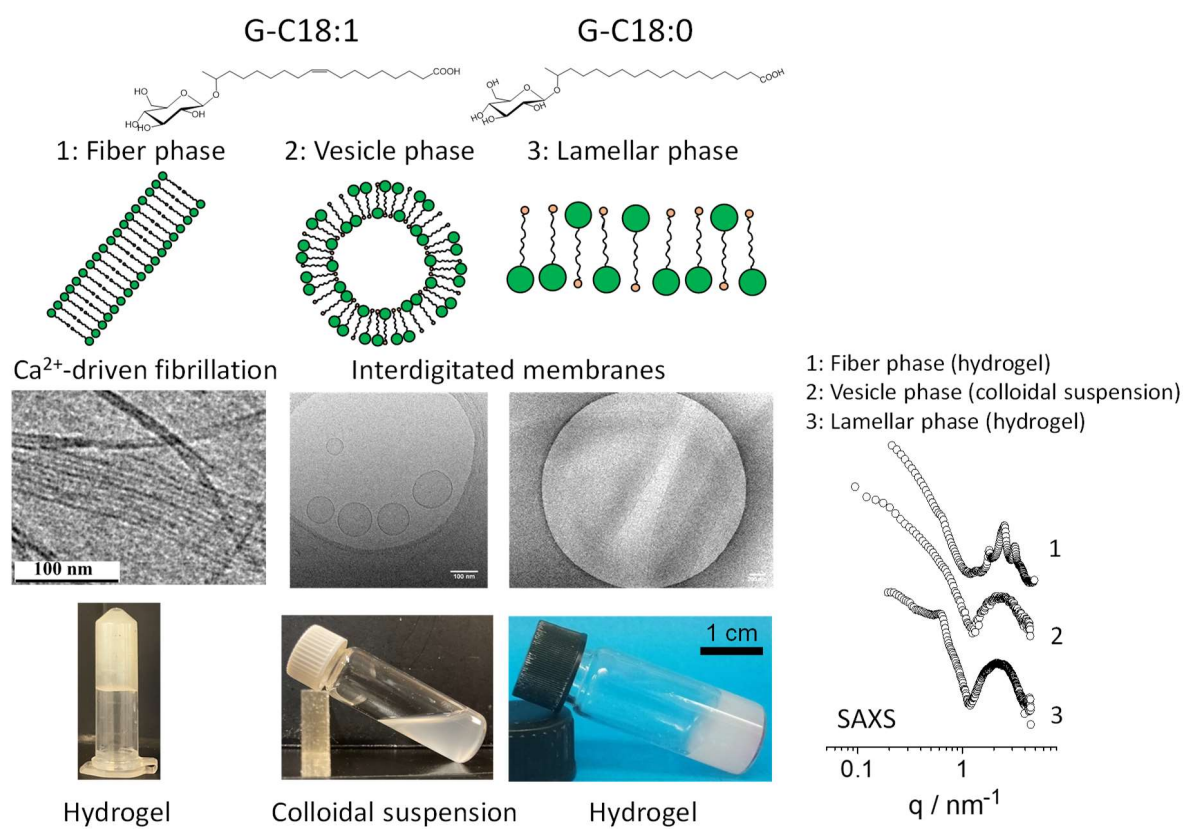

**Figure 1 – Chemical structure of monounsaturated, G-C18:1, and saturated, G-C18:0, bolaform glucolipids. Pictures, cryo-TEM images and SAXS profiles correspond to their Fiber phase hydrogel (*1*, G-C18:1),**[19] **Vesicle colloidal suspension (*2*, G-C18:1)**[18] **and Lamellar phase hydrogel (*3*, G-C18:0).**[22]

**Experimental Section**

*Chemical*. Glycolipid biosurfactant, glucolipid G-C18:1, was purchased from Amphistar (Gent, Belgium) and produced by the Bio Base Europe Pilot Plant (Gent, Belgium), lot N° APS F06/F07, Inv96/98/99 and used as such. The monounsaturated glucolipid G-C18:1 ($M_w$ = 460 g.mol$^{-1}$) contains a β-D-glucose unit covalently linked to oleic acid. The molecule is obtained by fermentation from the yeast *Starmerella bombicola ΔugtB1* according to the protocol given elsewhere.[37] According to the specification sheet provided by the producer, the batch (99.4% dry matter) is composed of 99.5% of G-C18:1, probed by HPLC-ELSD chromatography data. NMR analysis of the same compound (different batch) was performed elsewhere.[18] Glucolipid G-C18:0 ($M_w$ = 462.6 g.mol$^{-1}$) was obtained by catalytic hydrogenation reaction of G-C18:1. Hydrogenation is performed in an H-Cube Pro continuous-flow hydrogenation system (ThalesNano, Nanotechnology Inc.) equipped with CatCart® (height: 30 mm, Ø 5 mm) tubular reactors containing 10% Pd/C as catalyst. The system was washed with methanol, the reaction temperature and pressure were adjusted to 50 °C and 20 bar respectively and the ratio gas to liquid was fixed to 50%. Washing with MeOH was continued until steady state was reached. The stock solution of glucolipid G-C18:1 was prepared by dissolving 3.00 g (6.5 mmol) of G-C18:1 in 300 ml of MeOH resulting in the concentration of the substrate of 0.02 mol/L. In a typical experiment, the stock solution was loaded into the tubular reactor by an HPLC pump at 1 mL/min flow rate. After transfer of the entire stock solution, glucolipid G-C18:0 is obtained by evaporation of the solvent without further purification. Successful hydrogenation is easily confirmed by the disappearance of the olefinic protons in position $C_{9,10}$ ($H_{9,10}$, δ= 5.36 ppm in $^1$H NMR) of the aliphatic chain of G-C18:1, as shown elsewhere.[18] Deuterated water, $D_2O$, and DCl 37% are purchased from Sigma-Aldrich.

*Preparation of the samples.*

To reduce the incoherent neutron scattering from hydrogen and to maximize contrast, all samples were prepared in deuterated water, $D_2O$. The concentration of hydronium ions in $D_2O$ (pD) is adjusted by using DCl and NaOD solutions, prepared by diluting concentrated DCl (37 wt%) or NaOH pellets, in $D_2O$ to obtain DCl and NaOD solutions of 5 M, 1 M, 0.5 M and 0.1 M.

*Lamellar hydrogels (LG) from G-C18:0.* LG can be prepared at room temperature in a broad range of concentrations, from 5 mg/mL to 100 mg/mL, in the pH range between about 5 and 7 according to a protocol described in previous works.[22,23] In short, the G-C18:0 powder is dispersed in water at 25°C (C= 20 mg/mL, 43.3 mM, V= 1 mL) followed by sonication, adjustment of pD to 6.8 and sonication again. pD is adjusted with µL-amount (4 to 8 µL) of 5 M NaOD and refined with 1 M NaOD (10 to 15 µL), for a cumulative $[Na^+]_{OD}$ of about 50 mM. Further sodium ion can be added from the corresponding NaCl stock solution ( [NaCl]= 5 M, prepared in D$_2$O) so to achieve the total Na$^+$ concentration, calculated as follows: $[Na^+]_{total}$= $[Na^+]_{OD}$ + $[Na^+]_{NaCl}$, with $[Na^+]_{NaCl}$ being the concentration of Na$^+$ derived from the complementary NaCl solution. When Ca$^{2+}$ ions are added, we employ a 1 M CaCl$_2$ stock solution (in D$_2$O), added in µL amount. In the latter, we label the sample by the concentration of Ca$^{2+}$.

*Fibrillar hydrogels (FG) from G-C18:1.* FG are prepared at room temperature from G-C18:1 solutions at pH> 7 in a broad range of concentrations, between 5 mg/mL and 100 mg/mL, by adding a source of Ca$^{2+}$ or Ag$^+$ ions. More details on the synthesis, characterization and structure of G-C18:1 fibrillar gels are given in Ref. [19,21].

In this work, G-C18:1 is dispersed in D$_2$O at 20 mg/mL at 25°C and the pH is adjusted to about 7.5 by an initial addition of NaOD (5 M), followed by a refinement with few µL of more diluted NaOD (1 M, 0.5 M or 0.1 M). To trigger the sol-to-gel transition, a solution of CaCl$_2$ (1 M, in D$_2$O) is added so to reach a molar ratio [CaCl$_2$]/ [G-C18:1] of about 0.6. After the addition, the solution is stirred during about 30 s. According to *in situ* SAXS data, the micelle-to-fiber transition is immediate,[38] while according to rheology, hydrogels form within about one hour and their strength increases over time, within a span of few hours.[24] For the present work, considering the fact that the NSE experiment runs over several hours, the criterion of the hydrogel strength was not taken into account. Furthermore, despite the fact that it was not measured, one can still reasonably assume that the length scale probed by NSE (< 20 nm) is much smaller than the mesh size of the hydrogel.

*Vesicle suspension (VS) from G-C18:1.* According to previous data, vesicles start to form at room temperature below pH 7, around pH 6.2 and below in the concentration range between 5 mg/mL to 50 mg/mL.[17,18] Here, vesicles are then obtained by dispersing G-C18:1 in D$_2$O at 25°C at the concentration of 20 mg/mL and by adjusting pD to about 6 by means of µL-amount of concentrated DCl (5 M), followed by a refinement with few µL of more diluted DCl solution (1 M, 0.5 M or 0.1 M).

*Neutron spin-echo (NSE) experiments.* Neutron Spin-Echo (NSE)[39] measurements have been performed at the instrument IN15[40] at the Institut Laue-Langevin (ILL) in Grenoble (France). Four different wavelengths (λ) have been used, namely 13.5, 12, 10 and 8 Å allowing to reach maximum Fourier times $t = \frac{\gamma_N J_i m_N^2}{2\pi h^2}\lambda^3$ of 477, 335, 194 and 99 ns, respectively, where $\gamma_N$ and $m_N$ are the neutron's gyromagnetic ratio and mass, $J_i$ is the instrument's field integral and $h$ is Planck's constant. At the same time, we are covering a $q$-range from 0.03 to 0.14 Å$^{-1}$, where $q = \frac{4\pi}{\lambda} sin\left(\frac{\theta}{2}\right)$ is the modulus of the scattering vector, with scattering angle $\theta$. The data were corrected for resolution effects using graphite and the scattering from the aqueous background was subtracted.

To analyze the data, the Zilman-Granek (ZG) model[41] was applied. Starting from a Helfrich bending Hamiltonian,[42] the ZG model predicts a stretched exponential shape of the intermediate scattering function $S(q,t)$ (Eq. 1) with a stretch exponent $\beta = 2/3$, $I(q,t)$ being the spin-echo intensity at a given value of $q$ and time, $t$

$$\frac{I(q,t)}{I(q,0)} = S(q,t) = e^{-(\Gamma_{ZG}t)^\beta} \qquad \text{Eq. 1}$$

where

$$\Gamma_{ZG} = \alpha\gamma\left(\frac{k_bT}{k}\right)^{\frac{1}{2}}\left(\frac{k_bT}{\eta}\right)q^3 \qquad \text{Eq. 2}$$

from which, the scaled bending rigidity, *k*, is

$$\frac{k}{k_bT} = \left(\alpha\gamma\frac{k_bT}{\eta}\frac{q^3}{\Gamma_{ZG}}\right)^2 \qquad \text{Eq. 3}$$

where, $\alpha = 0.0069$ is a prefactor (commented below), $\gamma \approx 1$ for $\frac{k_bT}{k} \ll 1$, $\eta$ is the solvent viscosity (here taken as the viscosity of deuterated water, $\eta = 0.00109$ Pa.s at 25°C and $\eta = 4.0 \cdot 10^{-4}$ Pa.s at 70°C) [43], $k_b$ is the Boltzmann constant, $T$ is the temperature in Kelvin.

The relaxation mode observed in the $q$ and $t$ range of NSE is not a pure bending mode[44] but a combined bending-stretching mode, which also depends on the compressibility modulus $k_c$. $k_c$ is proportional to the bending rigidity[45] and results in a renormalized bending rigidity, which can simply be used in the framework of the Zilman-Granek model,[46] thus resulting in a modification of the prefactor, $\alpha$, in Eq. 2. The exact value of $\alpha$ is still a matter of debate, but the current consensus seems to point towards a value of $\alpha$ = 0.0069, which is different from 0.025 as originally suggested and it accounts for the fact that at the time and length scale of NSE a combined bending and stretching mode is observed.[44–48] The current consensus around the value of 0.0069 is also the result of matching the bending rigidity obtained by NSE (unrelaxed membrane rigidity), and pipette aspiration (bare bending rigidity) measurements.[49][50] In a recent review, Gupta et al.[31] give a comprehensive overview of the prefactors that were used by different groups.

At small $q$, de Gennes narrowing[51] causes a decrease in the otherwise constant $\frac{\Gamma_{ZG}}{q^3}$ in the $q$ range (~ 0.4 < $q$ / nm$^{-1}$ < ~0.8), where a structure peak is observed in the small angle neutron scattering pattern.[22] Therefore, we limited our analysis to $q$ > 0.8 Å$^{-1}$, where the static structure factor S(q)~ 1. This aspect will be discussed in more detail in the main text. Finally, the scaled $k$ for all samples is obtained by averaging the values of $\frac{k}{k_bT}(q)$ above 0.8 nm$^{-1}$, that is outside of the de Gennes narrowing range in $q$.

**Results and discussion**

The bolaamphiphilic nature of the microbial glucolipids in Figure 1 drives the formation of interpenetrated membranes,[17,18] rather than classical bilayers, while the intrinsic presence of a carboxylic acid group in the molecule generates an unpredictable distribution of carboxylic/carboxylate groups in the membrane itself.[22,23] Furthermore, the carboxylate can complex specific cations, which drive the rearrangement of the molecule into a crystalline fibrillar network. This behavior, published elsewhere, is shortly reviewed in the next paragraph.

At room temperature, for a pH range between 5 and about 7 in the absence of specific cations, the morphology of the self-assembled glycolipids strongly depends on the nature of the aliphatic chain: an unsaturated C=C bond drives vesicles as the stable phase, while a fully saturated chain drives the formation of flat membranes (Figure 1).[18] In both cases, the structures are colloidally-stable, most likely due to the presence of negative charges associated to the carboxylate form of the glucolipid. Interdigitated saturated glucolipid lamellae form viscous solutions, which, upon addition of an extra source of mono- or divalent counterion, unexpectedly, and immediately, undergoes a sol-to-gel transition.[22] Lipid lamellar hydrogels are relatively rare fluids, classically obtained from lamellar lipid bilayers modified with polymeric clusters.[52] However, lamellar structures of saturated glucolipid characterized by a macroscopic elasticity are easily obtained in water above 1 wt% and in a pH range from about 5 to 8. Similarly to other lamellar hydrogels reported in the literature, the lamellar order is driven by medium-range repulsive electrostatic interactions while the rheological properties are probably driven by the highly defective nature of the lamellar phase. However, defects, and consequent macroscale elasticity, are not enhanced and stabilized by polymeric inclusions[52,53] or charges,[54] as for classical lamellar hydrogels, but rather by medium/high ionic strength (10-50 mM for $Ca^{2+}$ and 50-500 mM for $Na^+$).[22] Finally, adding a source of calcium to the monounsaturated, C=C, glucolipid in its micellar phase drives a sol-to-gel transition and the gel structure of which is fibrillar.[19,21] This behavior is also unexpected, as fibrillar hydrogels are generally not observed for those molecules having a tendency to form membranes, but rather for a specific class of molecules, addressed to as low-molecular weight gelators, and often containing cyclic rings. Figure 1 shows the typical cryo-TEM and SAXS fingerprints, as well as the macroscopic look, of G-C18:1 fibrillar gels (1),[18,22] G-C18:1 vesicle suspensions (2)[18] and G-C18:0 lamellar gels (3).[19,21] The SAXS profile of (2) and (3) are typical of flat membranes, with the exception that the loose lamellar order in (3) is associated to the broad structure peak at q= 0.64 $nm^{-1}$. The SAXS profile of (1) is characterized by a larger number of structure peaks starting at *q*-values as low as 0.65 $nm^{-1}$ and identifying both the fibers' internal structure as well as their long-range association. The structural complexity of both the G-C18:0 lamellar and G-C18:1 fibrillar hydrogels are reported in detail elsewhere.[19,22–24]

To correlate the nature of the molecule, the structure of its self-assembled form, but also the effect of added salt on the macroscale properties, we employ NSE as a tool to probe the local membrane dynamics and rigidity. Figure 2 shows the time-dependent evolution of the intermediate scattering function for a fiber gel (FG), a vesicle suspension (VS) and a lamellar hydrogel (LG), while Figure S 1 reports similar experiments performed under different physicochemical conditions to probe the effect of temperature, ionic strength or type of counterion ($Na^+$ and $Ca^{2+}$) on lamellar gels. The vesicle suspension displays a much more rapid decay of $S(q,t)$ compared to both the fibrillar and lamellar gels, the slow decay of which, especially at q< 0.74 $nm^{-1}$, suggests a rigid membrane at length scales above about 8 nm. Interestingly, the decay of the LG strongly improves upon heating to 70°C (Figure S 1), being actually comparable with the data of the vesicle suspension, and showing a softening of the membrane and faster dynamics. This behavior may not come unexpected, as it was shown before that temperature softens the lamellar gel,[23] inducing a transition towards a vesicle phase.[18] Finally, adding $Na^+$ or $Ca^{2+}$ ions, on the contrary, does not sensibly change the profile of the intermediate scattering function (Figure S 1).

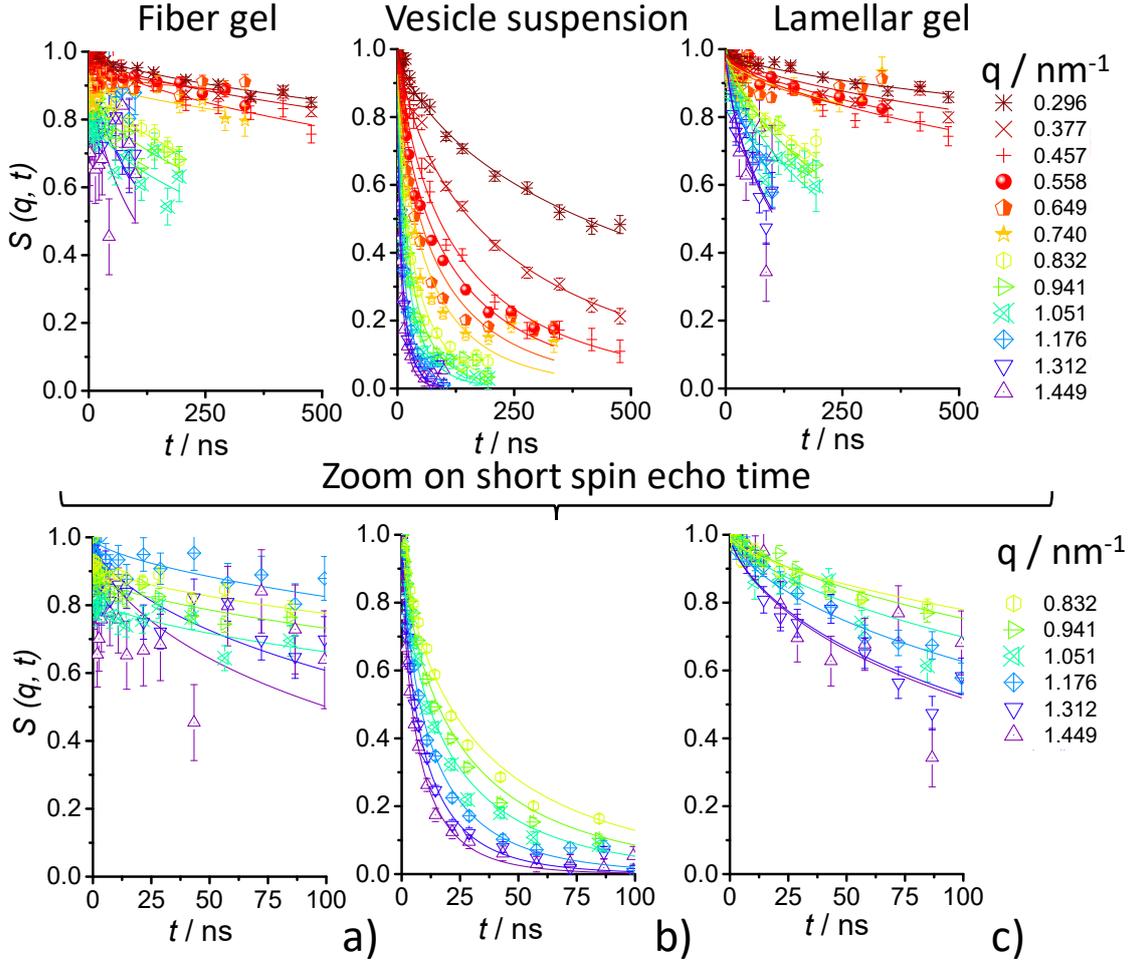

**Figure 2** –Normalized, *q*-dependent, spin-echo intermediate scattering function (full time scale on top, zoom up to 100 ns on bottom) for a) Fiber gels (G-C18:1), b) Vesicle suspension (G-C18:1) and c) Lamellar gels (G-C18:0) recorded at room temperature (25°C). Data are fitted with Eq. 1.

The fits of the intermediate scattering function using Eq. 1 provide the wavevector independent $\frac{\Gamma_{ZG}}{q^3}$ quantity, which can be plotted against *q*. The entire set of $\frac{\Gamma_{ZG}}{q^3}(q)$ profiles for selected samples explored in this work are presented in the *q*-range between 0.8 nm$^{-1}$ and 1.5 nm$^{-1}$ (Figure 3a,b), showing a *q*-independent behavior, as expected from Eq. 2, while a clear-cut, unexpected, *q*-dependency occurs for most samples in the *q*-range between 0.4 nm$^{-1}$ and 0.8 nm$^{-1}$, where $\frac{\Gamma_{ZG}}{q^3}(q)$ becomes, in the best case scenario, small and, in the worst case, it cannot be estimated at all. The full $\frac{\Gamma_{ZG}}{q^3}(q)$ datasets are shown in Figure S 2, which nicely shows the drop located at approximately *q*= 0.65 nm$^{-1}$ for most samples, except for the vesicle suspension sample, for which $\frac{\Gamma_{ZG}}{q^3}(q)$ is essentially homogeneous across the entire *q*-range.

The present dependency of $\frac{\Gamma_{ZG}}{q^3}$ with *q*, in apparent contradiction with Eq. 2, is actually explained by the so-called de Gennes narrowing[51] and it is related to the coexistence of a non-unitary, S(q)≠ 1, structure factor in the static neutron scattering function. This is illustrated in Figure S 2c-e, where the typical SAXS $I(q)$ profiles (left axis) for the lamellar, fibrillar gels and vesicles suspension are superimposed to the $\frac{\Gamma_{ZG}}{q^3}$ plot (right axis) against the same $q$-dimension. Figure S 2c,e nicely show how the drop in $\frac{\Gamma_{ZG}}{q^3}$ is practically superimposable to the broad low-$q$ peak for both the lamellar (Figure S 2e) and fiber (Figure S 2c) gels. This peak, despite the similar position in terms of $q$-value, identifies two different structures. In the lamellar gel, it was attributed to a loose lamellar stacking generated by long-range (> 20 nm) electrostatic repulsive interactions,[22] while in the fiber gel, it was attributed to the periodicity of side-by-side stacking of crystalline fibers.[19] Further proof that drop in $\frac{\Gamma_{ZG}}{q^3}$ is related to a structure factor is given by the independency of $\frac{\Gamma_{ZG}}{q^3}$ in the entire $q$-range explored in this work for the vesicles suspension (Figure S 2d), known to be essentially composed of small unilamellar vesicles, the SAXS signal of which is then characterized by a form factor only.[18] Considering the de Gennes narrowing effect in most of the present samples and the fact that the scaled bending rigidity is directly calculated from $\frac{\Gamma_{ZG}}{q^3}$ (Eq. 3), we assume that S(q)≠ 1 for 0.4 < $q$ / nm$^{-1}$ < 0.8 and S(q)→ 1 for $q$ > 0.8 Å$^{-1}$. For this reason, only the $\frac{\Gamma_{ZG}}{q^3}(q)$ for $q$ > 0.8 nm$^{-1}$ is considered to calculate the bending rigidity, although fluctuations in $\frac{\Gamma_{ZG}}{q^3}(q)$ out of the error bar could still perturb the quantification of the scaled bending rigidity, especially at high ionic strength ( [Na$^+$]= 750 mM) and $q$ < 1.2 nm$^{-1}$. This can be understood by looking at the typical SAXS and SANS spectra: the low-$q$ structural lamellar peak shifts with the ionic strength,[22] as classically expected for lamellar systems.[55]

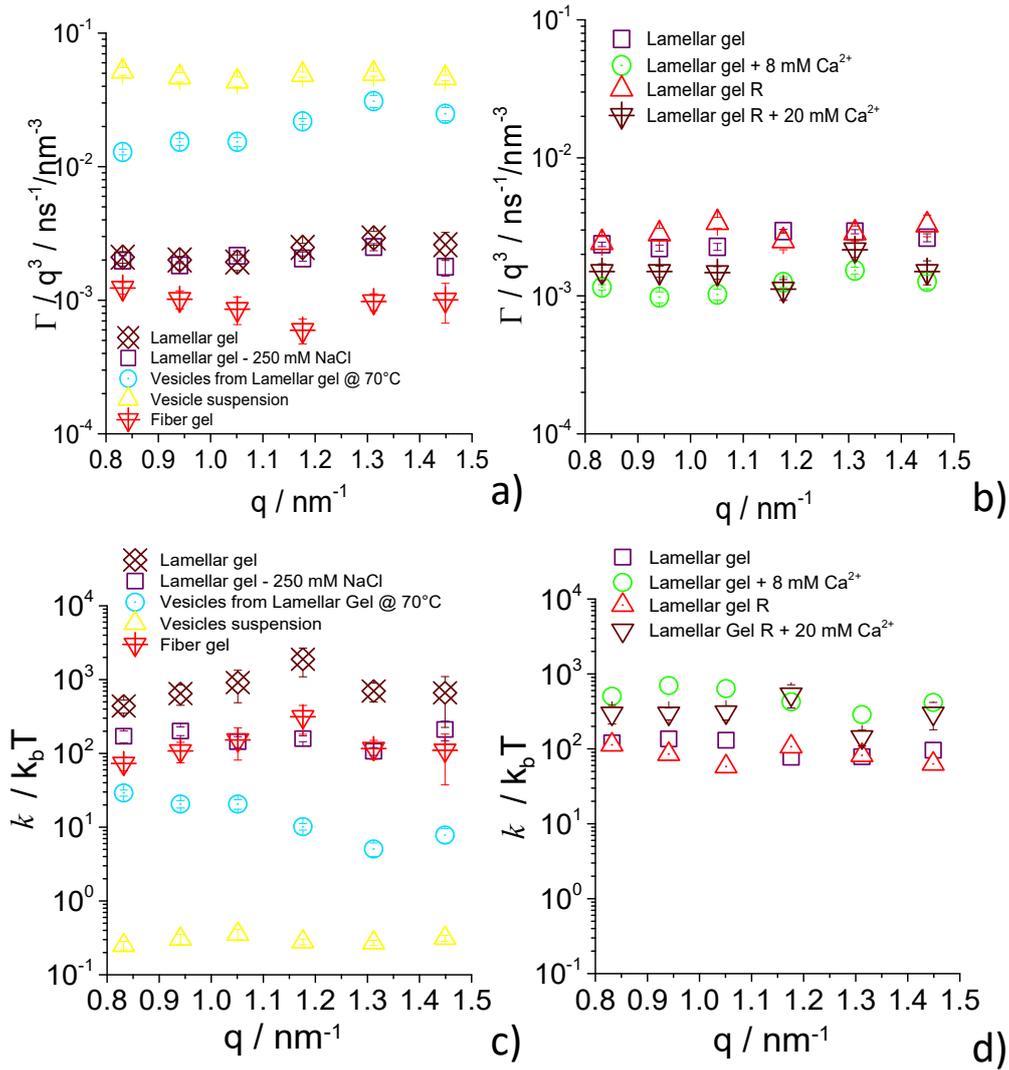

**Figure 3 –** a-b) Evolution of the Zilman-Granek decay parameter and c-d) of the scaled bending rigidity, $\frac{k}{k_bT}$, as a function of the wavevector, $q$, for various G-C18:1 (Vesicles suspension, Fiber gel) and G-C18:0 (Lamellar gels) self-assembled samples. Sample labelled "Vesicles from Lamellar Gel @ 70°C" corresponds to "Lamellar Gel – 250 mM NaCl" LG+ 250 Na$^+$" heated at 70°C. The lamellar-to-vesicle transition for lamellar gels was reported elsewhere.[18] The "R" in d) stands for "Replicate" and it refers to a freshly-made new sample.

The scaled bending rigidity, $\frac{k}{k_bT}$, can be derived from $\frac{\Gamma_{ZG}}{q^3}$ according to Eq. 3 with a prefactor $\alpha$= 0.0069, the choice of which is discussed in the Experimental section. The $q$-invariant evolution of $\frac{k}{k_bT}$ at $q$> 0.8 nm$^{-1}$ for most samples is shown in Figure 3c,d, with Figure S 3 showing the experiment performed in triplicate for the Lamellar gel (R and R1 in Figure S 3 indicate Replicate) and displaying a good reproducibility of the results. The entire $q$-range, displaying the de Gennes narrowing effect (sudden increase in $\frac{k}{k_bT}$ between $q=0.4$ nm$^{-1}$ and $q=0.8$ nm$^{-1}$) is on the contrary shown in Figure S 4. The global trend of $\frac{k}{k_bT}$ for all samples studied in this work is shown in Figure 4.

In terms of absolute values, Fiber gels display the strongest apparent bending rigidity, in the order of $10^3$ $k_bT$, while the lowest bending rigidity belongs to the Vesicle suspension, $k= 0.3 \pm 0.1$ $k_bT$ (Figure 4a), the radius of which lies in the order of hundred on nanometers. The bending rigidity of Lamellar gels depends on the type and amount of cation. The as-prepared LG, containing about 50 mM Na$^+$, has $k= 130 \pm 40$ $k_bT$, while adding increasing amount of Na$^+$ raises the apparent value of $k$ up to about 600 $k_bT$, with a certain linearity between 250 mM, 500 mM and 750 mM (Figure 4a). The same net increase of apparent bending rigidity of Lamellar gel is also achieved by adding a reduced content of a Ca$^{2+}$ source. Two different Lamellar gels of equivalent $\frac{k}{k_bT}$ within the error ($k$ ~100 $k_bT$, Figure 4b) undergo a multifold increase in $k$ (apparent $k$ ranges between $300 \pm 100$ $k_bT$ at 8 mM and $500 \pm 200$ $k_bT$ at 20 mM) when less than 20 mM Ca$^{2+}$ are added to the gel. Interestingly, heating a Lamellar gel at 70°C strongly reduces the bending rigidity, which drops to $k= 16 \pm 9$ $k_bT$.

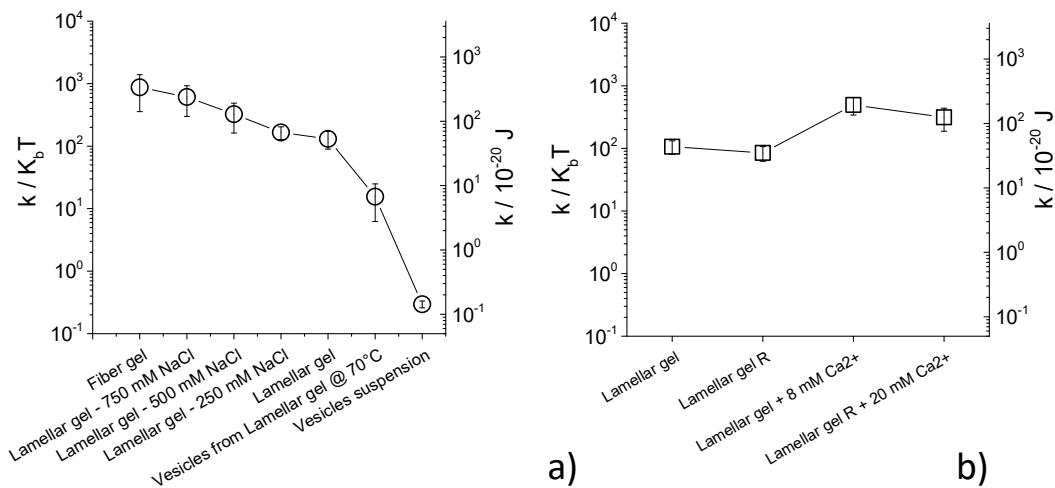

**Figure 4 – Scaled bending rigidity values measured for the set of samples studied in this work. R in b) stands for Replicate. Right-hand axis is rescaled in Joules for convenience of comparison with literature.**

**Discussion**

Bending rigidity is usually reported for lipid membranes in their fluid $L_\alpha$ and gel $L_\beta$ phase. The values of bending rigidity are commonly reported for lipidic liposomes and they fall in a broad range of values, ranging from less $< k_bT$ to more than 100 $k_bT$, depending on a number of conditions, like type of phospholipid, formulation with other lipids or sterols, pH, temperature, buffer, etc...[2,31,56] However, lipid membranes with bending rigidity in the order of 100 $k_bT$ are rare and generally observed for gel $L_\beta$ phase formed by saturated lipids at T≲ $T_m$, while $k$ generally varies between ~10 and ~60 $k_bT$ for $T>T_m$.[2,31] Bending rigidity for swollen lamellar phases or lamellar gels are in the order of $k_bT$ or less[57–59] (Table 1), while for free-standing lamellar membranes, values in the order of 5 $k_bT$ (~$2.10^{-20}$ J, assuming room temperature) were reported.[60] In the case of self-assembled fibers, bending rigidity depends on the system and it is generally expressed in units of $l_p k_b T$, with $l_p$ being the persistence length as the unit of the bending energy is [energy/length] as opposed to planar systems where the unit of the bending energy is [energy/area]. In principle, when observing bending fluctuations in a one dimensional system, $S(q,t)$ should show a stretched exponential relaxation with a stretching coefficient of 3/4 as opposed to 2/3, as found in planar systems.[41,61] Here, we do not observe such a change in the shape of the intermediate scattering function and therefore treat all data as coming from a quasi-planar system. This seems justified not only on a phenomenological level. The long dimension of a single fiber can be considered as "infinite", while its cross-section has one dimension of about 10 ± 1 nm and the other being in the order of the molecular size.[19] In addition, fibers associate side-by-side into planar rafts as wide as 100 nm,[19] so that overall the fibers can be treated as planar on the relatively short length scales of NSE. For the purpose of comparison with literature values for other fibrillar systems we adopt an *ad hoc* rescaling, where we simply multiply the obtained planar bending rigidity in units of [energy] by the typical width of the fiber to obtain a value in [energy*length] which should at least allow for a rough qualitative comparison.

To the best of our knowledge, the use of NSE to determine bending rigidity of self-assembled fiber is extremely rare, if not unique,[62] while microscopy methods (AFM, fluorescence microscopy)[63–69] combined with sonication[70–72] or numerical modelling[73–75] are more common, as they provide directly access to $l_p$. Most data were generally recorded for self-assembled peptides; for instance, MAX1 and MAX8 peptides were reported to have $l_p$ of 55 nm, providing a bending rigidity in the order of $2·10^{-28}$ Nm$^2$,[62] while polymerized cyclic peptides had shown bending rigidity two orders of magnitude higher, $4·10^{-26}$ Nm$^2$ (Table 1).[76] However, literature is rich of data and persistent length values vary from tens of nm to the micron scale, with corresponding bending rigidity values ranging from $10^{-28}$ Nm$^2$ to $10^{-28}$ Nm$^2$.[63,64]

In the following, each system presented in this work will be specifically discussed.

**Table 1 – Typical range of bending rigidity values reported for various classes of soft self-assembled systems. *: range of values estimated from values of persistence length spanning from 50 nm to 10 μm.**

| Structure | Bending rigidity | Reference |
|---|---|---|
| Vesicles | $k_bT$ to ~100 $k_bT$ | [2,31] |
| Lamellar gels | ~ $k_bT$ | [57] |
| Swollen lamellar phases | ~ $k_bT$ | [58,59] |
| Standing planar membranes | < 5 $k_bT$ | [60] |
| Self-assembled fibers | $10^{-28}$ to $10^{-26}$ Nm$^2$ [*] | [62–64,76] |
| Vesicle suspension | $0.30 \pm 0.04$ $k_bT$ | This work |
| Lamellar gel | $130 \pm 40$ $k_bT$ | This work |
| Lamellar gel + ion | 130 $k_bT$ to 600 $k_bT$ | This work |
| Fiber gel | $900 \pm 500$ $k_bT$ | This work |

*Lamellar gels*. In this work, Lamellar gels with [Na$^+$] concentration below 250 mM show $k$ in the range of 100 $k_bT$. Lamellar gel of G-C18:0 were described as a network of defectuous free-standing, colloidally-stable, interdigitated G-C18:0 membranes[18,22] characterized by a fairly rigid $P_{β'}$ phase, with the $T_m$ of G-C18:0 being above room temperature.[22] Literature values of bending rigidity for free standing lamellar membranes (DOPC), a morphology comparable to the one of G-C18:0 Lamellar gels, are almost two orders of magnitude lower, even in the presence of calcium ions (< 5 $k_bT$).[60] Similar low values of the bending rigidity are generally assumed for swollen lamellar phases and lamellar gels (Table 1), the structure of which is similar to the G-C18:0 gels. On the other hand, values in the order or 100 $k_bT$ were reported for saturated phospholipid vesicles, possibly containing rigidifiers like sterols,[2,4,31,56] whereas G-C18:0 contains itself a stearic moiety. One of the few exceptions to the above seems to be the work of Seto et al.,[77] reporting exceedingly high values ($k$= 2.1·10$^{-17}$ J at 38.5°C, that is $k$ ~4900 $k_bT$) of bending rigidity for a swollen lamellar, $L_S$, phase composed of 14 wt% DPPC. They have also reported values in the order of $k$= 375 $k_bT$ (1.6·10$^{-18}$ J, at 36°C) for an interdigitated $L_{βI}$ phase. If the origin of such high values, in contrast with literature Table 1, was not discussed, it could be related to presence of lamellar crystals in their structure. Whether or not interdigitation could explain bending rigidity above 100 $k_bT$ seems to be excluded by Kelley et al.,[78] which showed that the bending rigidity of SMPC with a higher degree of interdigitation was estimated to be in the order 80 $k_bT$, after rescaling of the prefactor, as also demonstrated by the order of magnitude for $\Gamma/q^3$, ~ 5·10$^{-3}$ ns$^{-1}$/nm$^{-3}$. Overall, compared to the wide majority of literature data, neither the swollen flat structure nor interdigitation can explain the high values found for G-C18:0.

Lower values of $k$ for the Lamellar gel sample are only obtained by heating the sample at 70°C, above its $T_m$. The corresponding value of $k= 16 \pm 9$ $k_bT$ is in the range of more standard values reported before for liposomal systems.[2,31] The loss of a close-to-ten factor in the bending rigidity is explained by the higher fluidity of the lamellar phase, causing a softening of the hydrogel[23] and a possible lamellar-to-vesicle transition.[18] Loss in the bending rigidity with temperature is known since long time[1] and largely reported before for many lipidic systems,[2,79] especially when approaching the $T_m$.[2]

Overall, it seems reasonable to attribute the high bending rigidity of G-C18:0 Lamellar gels to its composition and not to its collectively flat morphology. However, this conclusion raises the following question. If flat lipid membranes, may them be in a swollen, possibly gelified, state, are classically characterized by low bending rigidity, why G-C18:0 is able to collectively assemble into rigid, colloidally-stable, flat membranes? This issue shows that the exceedingly high values of bending rigidity found for the G-C18:0 Lamellar gel may actually neither be explained by the flat morphology of the membranes (low $k$ expected) nor by their composition (saturated lipids are expected to form vesicles with stiff membranes). Salt-dependent experiments may help understanding this unique system.

Increasing the salt content at constant (room) temperature has a drastic influence on $k$, which increases between a factor 3 and up to almost 6, depending on the type of ion and concentration, with the effect of $Ca^{2+}$ being much more prominent than $Na^+$. Millimolar amounts of $Ca^{2+}$ produce a comparable increase in $k$ as 500 mM of $Na^+$ (Figure 4). At the same time, a monotonic increase of a factor 5.5 in $k$ is observed when going from as-prepared G-C18:0 Lamellar gels ( $[Na^+]$ ~50 mM) to $[Na^+]$= 750 mM. Values of the bending rigidity above 100 $k_bT$ are uncomparable with the literature and their actual physical sense is itself unclear. Raw data of the intermediate scattering function are noisy and error bars are large, meaning that the actual values are not accurate, as there is barely any decay of the intermediate scattering function within the experimental time window due to bending undulations. On the one side, values above well above 100 $k_bT$ should most likely be interpreted as representative of essentially rigid membranes. On the other side, increasing sodium content, or employing calcium ions, show a specific trend of increasing bending rigidity and actually corresponding to the qualitative improvement of the elasticity of the lamellar gels.[22]

Interpreting the correlation between added salt and bending rigidity of lipid lamellar phases is not straightforward, as contradictory data were reported in the literature and hardly at such a high salt concentration, as recently discussed by Dimova.[2] As a matter of fact, for microemulsions and lipid vesicles, it was found that introducing charges in the membrane itself has relatively little influence.[35,80] Dimova showed that bending rigidity of POPC and POPC:POPG membranes actually decreases with NaCl (limit of 45 mM explored),[81] but others have found that adding salt to otherwise neutral membranes result in a pseudo increase in bending rigidity,[9,82] but the increase was mild and quantified to less than 10 $k_bT$ between 0 and 470 mM NaCl. Buffer was also shown to have an impact on the bending rigidity of POPC membranes, but variation was only contained between 31 $k_bT$ and 41 $k_bT$.[83] Similarly, even the method of vesicle preparation (e.g., spontaneous swelling, electroformation, phase-transfer method, …) could be used as an argument to justify variation in $k$, but once again, the range of bending rigidity values for POPC giant vesicles was contained between 20 $k_bT$ and 37 $k_bT$.[3] In other words, to the best of our knowledge, values of $k$ of several hundred $k_bT$, specifically induced by adding salt, are simply not discussed in the current literature. However, the observations published in Ref. [11], which associate increase in $k$ to an increased multilamellar character of vesicles, could help understanding the current G-C18:0 Lamellar gel system.

In our previous work, we had proposed the hypothesis that salt generates defects in the G-C18:0 lamellar gel. However, the as-prepared lamellar gels themselves contained such a high density of defects (observed by polarized light microscopy) that the actual effect of adding salt did not allow a direct verification of the hypothesis. On the other hand, salt content close to 1 M outnumbers the amount of negative charges, in the mM range, meaning that charge screening and a dense lamellar precipitate should form at much lower salt content than what actually experimentally observed.[22] The high bending rigidity, demonstrated by NSE experiments, could then be associated to local nanometer-scaled lamellar domains, which act as nodes of the gel network. Excess of salt could then increase the content of lamellar domains characterized by, we speculate, a shorter $d$-spacing than in the bulk and overall resulting in an apparently stiffer gel. Given the complexity of this soft material, this hypothesis is hard to verify, but a series of SANS data collected at various G-C18:0 and salt concentrations[22] had shown the coexistence of two peaks, the one at higher $q$-values (noted by * in Ref. [22]) being sharper than the other, thus suggesting an improved local order. If the presence of two lamellar peaks at higher salt content could certainly be taken as proof of the coexistence of two lamellar networks, it is still debatable whether or not the second network could be associated to nodes of the lamellar gel.

*Vesicle suspension.* The Vesicle suspension prepared at room temperature from G-C18:1 shows the lowest value of the bending rigidity (~ 0.2 $k_bT$) compared to all phases studied in this work. Values below $k_bT$ are not uncommon,[56,84] although most of them are generally in the order of at least few $k_bT$.[2,31] Explanation for such low values is unclear and it could be related to the interdigitated structure of the membrane. In this regard, Jadidi et al.[85] calculated bending rigidity for interpenetrated membranes in the order of 7.6 $k_bT$ for the interdigitated phase, when temperature is in the order of 0°C (3.6·$10^{-21}$ J), while Kirch et al.[86] have shown, using numerical modelling as well, that DPPC close to its $T_m$ (320 K) has a higher degree of interdigitation and the corresponding bending rigidity, both measured and calculated, is rather in the order of 10·$10^{-20}$ J, that is about 23 $k_bT$. If the values reported in these studies then seem to show that, compared to a classical bilayer structure, interdigitation *per se* does not have a significant impact on the bending rigidity and cannot help explaining the low values found for G-C18:1 vesicles, one should still keep in mind the following. Interdigitated membranes, also known as monolayer lipid membranes (MLM) when prepared from bolaamphiphilic molecular systems, are long described for macrocyclic polyether lipids, analogues of, or extracted from, extremophile microorganisms.[25] Interestingly, the unique properties, like resistance to both low and high temperatures,[26] of MLM was explained by variations of the phase transitions temperatures, enthalpy and entropy change, when compared to their double tail phospholipid bilayer analogues.[26,28,29] Such variations were eventually connected to the monolayer structure but also to the chemical variability in the acyl and head groups of natural bolaamphiphiles.[26] Even if one should not make a straightforward correlation between the data reported on bolaamphiphiles and G-C18:1, it is certainly curious to note that a similar ease in obtaining G-C18:1 vesicles by simple sonication[17,18] was also reported for bolaform, hydroxylated or glycosylated, C24 macrocyclic diols (samples 16a, 16d, 16e in Ref. [25]). In this regard, even if the molecular understanding is still unclear, the low *k* measured for G-C18:1 vesicles are not that surprising and it could explain the spontaneous vesiculation of this molecule, as it was shown before that a flexible membrane should facilitate the formation of vesicles.[87]

The diameter of G-C18:1 vesicles is polydisperse and it can vary between a range of about one hundred nm to few microns.[18] The low bending rigidity could then also be explained by the contribution of the vesicle translational diffusion to the intermediate scattering function. To account for this hypothesis, the Vesicle suspension sample is fitted using the Zilman-Granek model (Eq. 1) multiplied by a translational diffusion term (see, for instance, Eq. 4 in Ref. [31]), shown in Figure S 5a, which also shows the typical fit for the intermediate scattering function recorded at $q= 1.449$ nm$^{-1}$ and using the modified ZG model. The comparison between the bending rigidity obtained for all $q$-values using the ZG model with and without the translational diffusion factor is shown in Figure S 5b, for a liposome of radius, $R= 50$ nm, a value settled in the smallest range known for G-C18:1 vesicles, according to cryo-TEM.[18] Overall, Figure S 5b shows that taking into account the diffusion term does not sensibly change the magnitude of the bending rigidity, the average of which over all $q$-values increases only by 30%, from $\bar{k}= 0.30 \pm 0.04$ $k_bT$ (ZG model only) to $\bar{k}= 0.40 \pm 0.07$ $k_bT$ (ZG with diffusion), but still sensibly below 1. Employing an even smaller, rather unrealistic for G-C18:1, vesicle radius of 25 nm would only increase $\bar{k}$ to about 0.5 $k_bT$. This suggests that the low values of $k$ cannot be explained by translational diffusion arguments.

Another explanation could be rationalized with a simple scaling argument. The bending rigidity of a fluid bilayer in principle scales as $t^2$, $t$ being the membrane thickness.[88] While the bilayer thickness of a non-interdigitated bilayer for C-18 lipids is on the order of 6 nm, for the systems investigated here it was previously found that the bilayer thickness is rather in the order of 3 nm, due to interdigitation,[17,18] which should lead to a bending rigidity significantly below typical values on the order of 20 to 30 $k_bT$, although still higher than the experimental values measured here.

*Fiber gels*. Fiber gels are characterized by the highest values of bending rigidity, in the order of 1000 $k_bT$. As these high values of $k$ result in a very limited decay of the intermediate scattering function, their reliability is rather limited and the curves could in principle be interpreted as having an elastic plateau. For the sake of comparability we nevertheless chose to describe them using eq. (1) and on a qualitative level, this is certainly valid. However, the exact values should be considered with care. Also, this value is hard to be compared with previous literature, as, to the best of our knowledge, within the frame of the study of the dynamics of proteins,[89] NSE was essentially applied to study the dynamics self-assembled peptide MAX1 and MAX8 fiber hydrogels,[62] or during fibrillation of amyloidogenic insulin,[90] but never to low-molecular weight amphiphiles. And even in the former cases, only the rigidity of MAX1 and MAX8 was actually evaluated. More specifically, peptide fibers were modeled as rigid cylinders and their corresponding $q$-dependent decay parameter followed a 8/3 power law, predicted for semiflexible chains. For both peptides, authors estimated the segmental diffusion coefficients, $D_G$, in the order of $10^{-2}$ nm$^{8/3}$ ns$^{-1}$, from which they did not calculate the bending rigidity, but a length scale (~ 3 nm) corresponding to a given number of inflexible hairpin units along the length of the fibril.

The G-C18:1 fibrillar gels prepared in this work are structurally different than the peptidic systems reported by Branco at al.. G-C18:1 fibers have a cross-section of approximately 10 nm in width and "infinite" length, but they also associate into two-dimensional rafts of width as large as about 100 nm. Such pseudo two-dimensional morphology at the length scale probed by X-rays features a global -2 exponent in the corresponding log(I)-log(q) scale in the corresponding SAXS profiles.[19] Locally, G-C18:1 fibrillar systems are then structurally closer to membranes than to unidimensional fibers. As a matter of fact, plot of the spin-echo intermediate scattering function against $t^{2/3}$ or $t^{3/4}$ (semiflexible chain)[91] for the G-C18:1 Fiber gel sample studied here does not show any conclusive behavior (Figure S 6). In this regard, the NSE data recorded for the G-C18:1 Fiber gel are not directly comparable to the ones recorded for MAX1 and MAX8 peptides. On the other hand, a qualitative comparison can be done between the spin-echo intermediate scattering functions of MAX1/MAX8 peptide fiber gels and G-C18:1 Fiber gel sample. In the comparable length-scale between 5 nm and 30 nm ($q$-range $0.2 < q$ / nm$^{-1}$ $< 1.25$) at spin-echo time below 50 ns, the intermediate scattering functions decay between 1 and 0.65 for the G-C18:1 Fiber gel, while it decays rather between 0.8 and 0.25 for MAX1/MAX8, the latter being more in the intermediate range for the vesicle suspension and lamellar gels found here.

On the basis of the present, quite limited, literature data based on NSE, one cannot directly compare the bending rigidities between self-assembled peptides and glycolipid amphiphiles fibers, it seems that the G-C18:1 fibers are actually stiffer. To confirm this assumption, it could be useful to use the bending rigidity data recorded on G-C18:1 Fiber gels and estimate the corresponding energy per meter (Nm$^2$), classically provided for self-assembled nanofiber systems. The bending rigidity measured here, in the order of 1000 $k_bT$, is equivalent to 4.1 10$^{-18}$ N·m. If one assumes a minimum width 10 nm (single fiber) and a maximum width of about 100 nm (raft),[19] one finds a range of energy per meter contained between 4.1·10$^{-26}$ Nm$^2$ and 4.1·10$^{-25}$ Nm$^2$, to be compared with literature values recorded on self-assembled peptide fibrils, often using microscopy.[62–64,76] In terms of equivalency, the G-C18:1 Fiber gel can be compared to peptide fibers characterized by a persistence lengths more than 10 μm, a range among the stiffest found in the literature and possibly explaining the interesting elastic properties of G-C18:1 gels and their resistance to stress, even at high temperatures.[19,24]

In summary, this work measures for the first time the bending rigidity of aqueous self-assembled structures, vesicles, lamellae and fibers, of two microbial glucolipids bioamphiphiles, also known as biosurfactants. It found that:

- Vesicular membranes of G-C18:1 and lamellar membranes of G-C18:0 are interdigitated (monolayer membranes) and have unusually small (< $k_bT$) and high (< 100 $k_bT$) $k$, respectively. As per comparison, the wide range of phospholipid-based bilayer membranes are characterized by $k$ in the order of few $k_bT$ and up to about 100 $k_bT$, generally in the presence of cholesterol and saturated lipids. The reason for such discrepancies is not clear at the moment, and it deserves further work. The bolaamphiphilic nature of the glucolipids and the interdigitation of the membrane could partially explain the present data, as monolayer membranes were demonstrated to have different physical properties (phase transition temperature, entropy and enthalpy but also fluidity) compared to bilayer membranes.[25,27–29] Other effects, conformational (in the sugar or acyl chain groups) or structural (e.g., lamellar ordering) should not be excluded.
- Fibers prepared by cross-linking G-C18:1 molecules with $Ca^{2+}$ ions display a rigidity comparable to peptide fibers.

**Conclusions**

This work quantifies for the first time the biophysical properties of three different self-assembled forms of new glycolipids obtained by microbial fermentation and known in the literature to belong to the family of biosurfactants. Neutron spin-echo spectroscopy ($q$-range between 0.3 nm$^{-1}$ - 21 nm - and 1.5 nm$^{-1}$ - 4.1 nm – and spin-echo time up to 500 ns) is used to probe the dynamics of vesicles suspension and fiber gels prepared from glucolipid G-C18:1 as well as lamellar gels prepared from glucolipid G-C18:0, both compounds being the only molecular component. It is found that the vesicle suspension has a bending rigidity value of $k$= 0.30 ± 0.04 $k_bT$, much below most biomembranes, while lamellar gels have $k$= 130 ± 40 $k_bT$. This value is in the upper most range of bending rigidities found in the literature and generally measured on gel phase, often containing cholesterol, while similar gels with similar lamellar structure rather show bending rigidity values in the order of $k_bT$. This discrepancy could be explained by both the saturated nature of G-C18:0 but it could also reveal an atypical gel structure, the network knots of which could be constituted by ordered lamellar domains, as observed by small angle X-ray scattering in corresponding structural studies previously published. Finally, fiber gels show $k$= 900 ± 500 $k_bT$, the physical meaning of which should more be associated to a persistence length. However, G-C18:1 fiber gels were shown to have a complex "nanofishnet" structure, where a fraction of the network knots could be constituted by fibrous lamellar rafts, the size of which is in the length scale probed by NSE. The high value of *k* could then be explained by such an atypical structure providing strong and stable hydrogels.

This work reveals two complementary aspects. From the fundamental perspective of membrane science, NSE shows that the elasticity of two fairly classical self-assembled flat and vesicular membranes prepared from bolaform glucolipid amphiphiles is profoundly different than what one expects on the basis of decades of research on phospholipids. Indeed, vesicles with bending rigidity being a fraction of $k_bT$ and cholesterol-free lamellae with bending rigidity above 100 $k_bT$ are rare, if not unique systems that deserve further studies to better understand the origin of such discrepancy. From a soft matter perspective, this work also shows that non-conventional biological glycolipids are able to assemble into fairly classical structures (vesicles, lamellae, fibers) characterized by unusual properties, which could probably be exploited further to tune the collective behavior of other classical amphiphiles, like surfactants, low-molecular weight gelators or membrane forming lipids.

**Acknowledgements**

We thank the ILL for the allocation of beam time and for access to the ILL computing cluster. The raw data of the NSE measurements are available under dx.doi.org/10.5291/ILL-DATA.9-13-915.

**Financial support**

SANS experiments have been supported by ILL, proposal N°9-13-778.

# Measuring the bending rigidity of microbial glucolipid (biosurfactant) bioamphiphile self-assembled structures by neutron spin-echo (NSE): interdigitated vesicles, lamellae and fibers

Niki Baccile,[a,*] Vincent Chaleix,[b] Ingo Hoffmann[c]

[a] Sorbonne Université, Centre National de la Recherche Scientifique, Laboratoire de Chimie de la Matière Condensée de Paris, LCMCP, F-75005 Paris, France
[b] Université de Limoges, Faculté des sciences et techniques, Laboratoire LABCiS - UR 22722, 87060 Limoges
[c] Institut Laue-Langevin, 38042 Grenoble, France

**\* Corresponding author:**
Dr. Niki Baccile
E-mail address: niki.baccile@sorbonne-universite.fr
Phone: +33 1 44 27 56 77

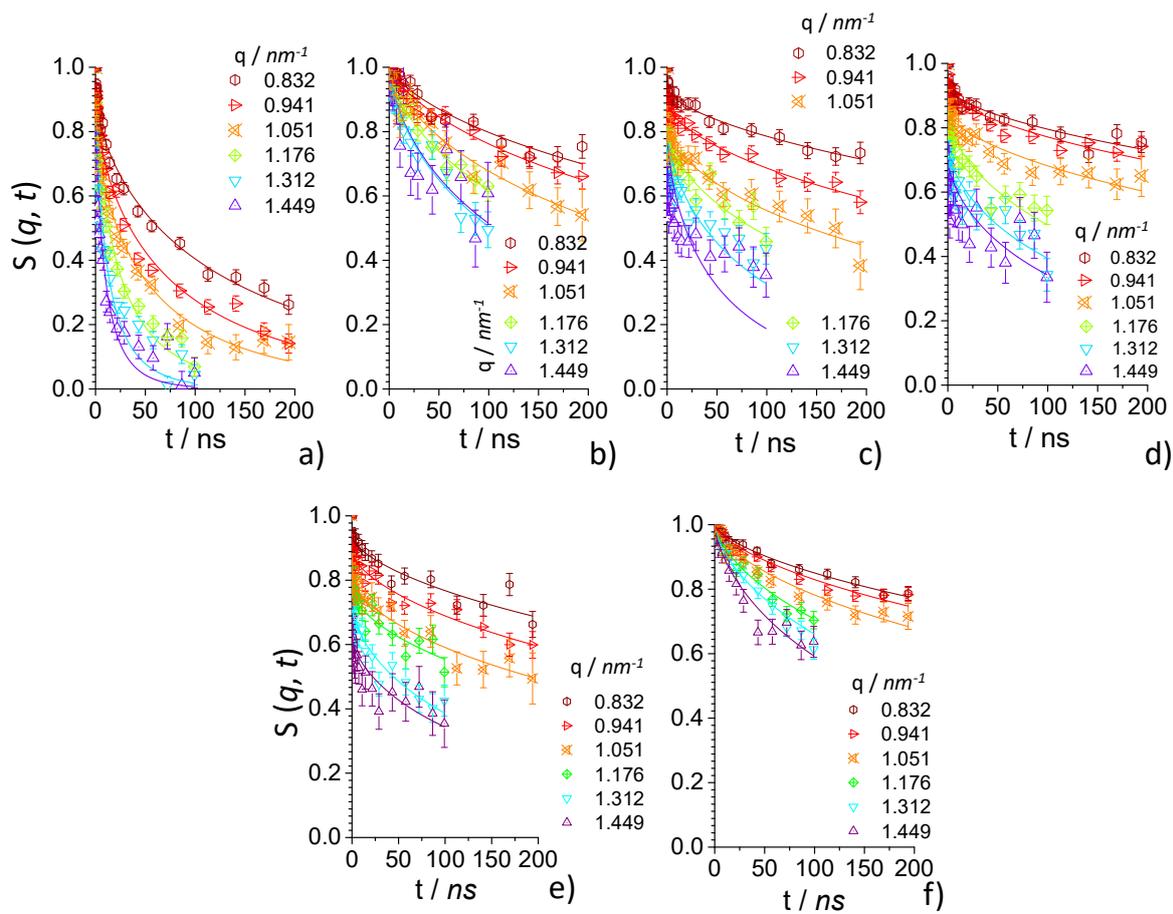

**Figure S 1 – Normalized, *q*-dependent, spin-echo intermediate scattering function a) Vesicles from Lamellar gels @ 70°C, b) Lamellar Gel – 250 mM NaCl, c) Lamellar Gel – 500 mM NaCl, d) Lamellar Gel – 750 mM NaCl, e) Lamellar Gel – 8 mM CaCl$_2$, f) Lamellar Gel  R – 20 mM CaCl$_2$.**

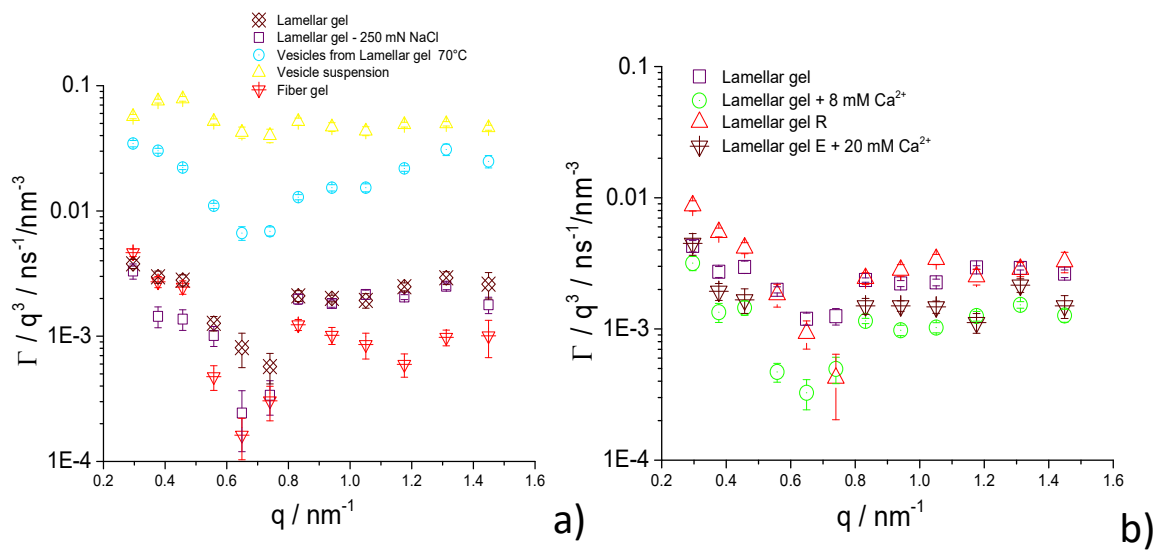

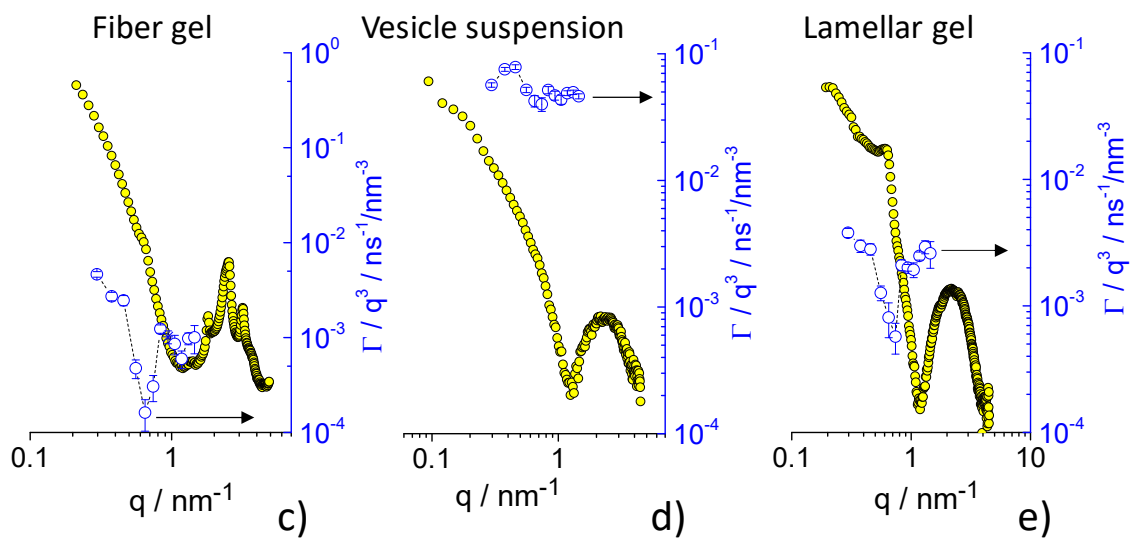

**Figure S 2 – a-b)** Evolution of the Zilman-Granek decay parameter as a function of the full-range wavevector, $q$, for various G-C18:1 (Vesicles suspension, Fiber gel) and G-C18:0 (Lamellar gels) self-assembled samples. Sample labelled "Vesicles from Lamellar Gel @ 70°C" corresponds to "Lamellar Gel – 250 mM NaCl" LG+ 250 Na$^+$" heated at 70°C. The "R" in b) stands for "Replicate" and it refers to a freshly-made new sample.

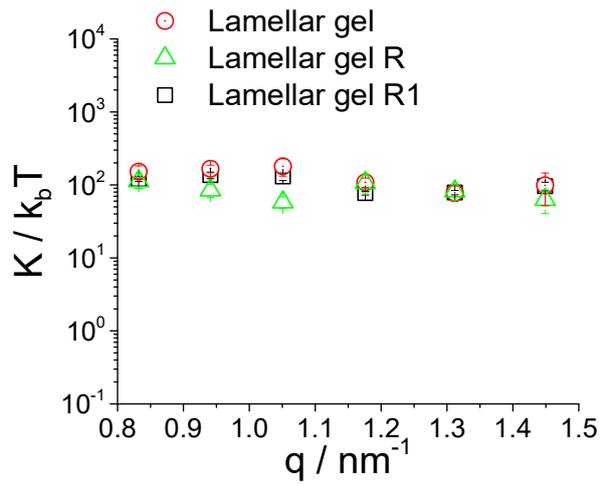

**Figure S 3** – Evolution of the scaled bending rigidity, $\frac{k}{k_bT}$, as a function of the wavevector, $q$, for three different Replicate of the G-C18:0 Lamellar gels.

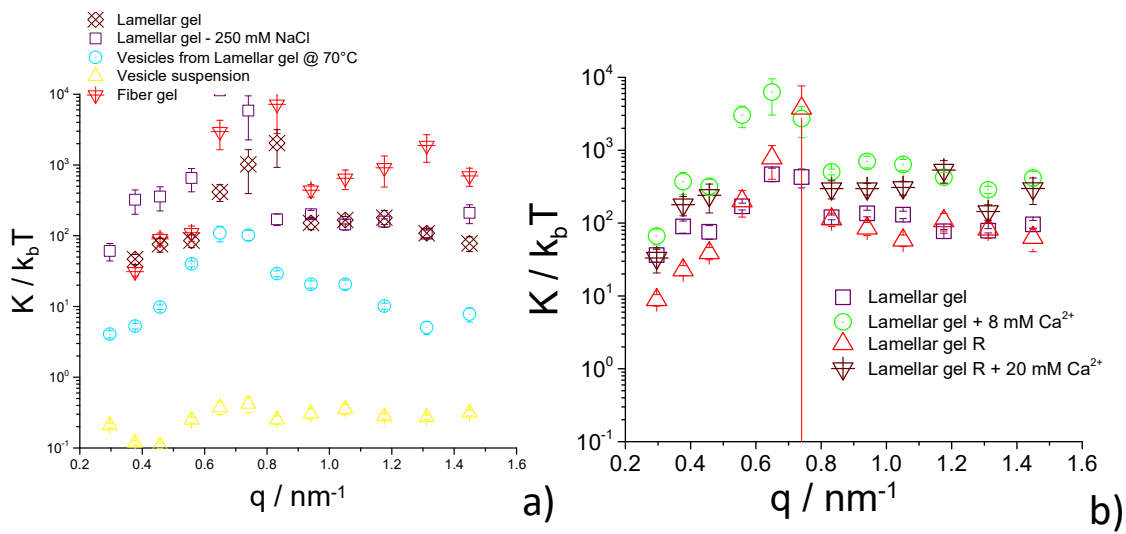

**Figure S 4 – Evolution of the scaled bending rigidity, $\frac{k}{k_bT}$, as a function of the full-range wavevector, *q*, for various G-C18:1 (Vesicles suspension, Fiber gel) and G-C18:0 (Lamellar gels) self-assembled samples. Sample labelled "Vesicles from Lamellar Gel @ 70°C" corresponds to "Lamellar Gel – 250 mM NaCl" LG+ 250 Na$^+$" heated at 70°C. The "R" in b) stands for "Replicate" and it refers to a freshly-made new sample.**

Vesicle suspension sample at q= 0.1449 nm$^{-1}$

Fit: Zilman-Granek model (ZG) with Diffusion term

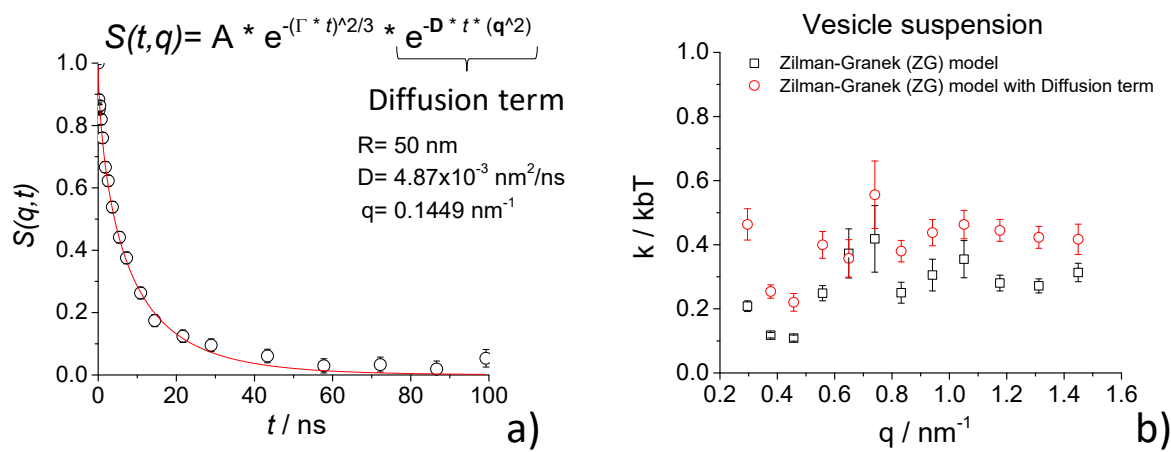

Figure S 5 – a) Plot of the spin-echo intermediate scattering function recorded for G-C18:1 Vesicle suspension sample (T= 25°C) as a function of spin-echo time. The fit is performed using the Zilman-Granek (ZG) model (Eq. 1 in the main text) to which a translational diffusion term, $e^{-Dtq^2}$, is included so to account for the diffusion of the vesicles. *D* is the diffusion coefficient (in nm$^2$/ns units) corresponding to an spherical object (here, a vesicle) of radius, *R*= 50 nm, diffusing in water. *D* is calculated with the classical Stokes-Einstein equation, $D= \frac{k_bT}{6\pi\eta R}$, with the dynamic viscosity of water, H$_2$O, $\eta$= 0.89x10$^{-3}$ Pa.s and T= 298 K (25°C). b) Dependency of the bending rigidity, *k*, with *q* for the Vesicle suspension sample calculated with the ZG (squares) and ZG with diffusion (circles) model.

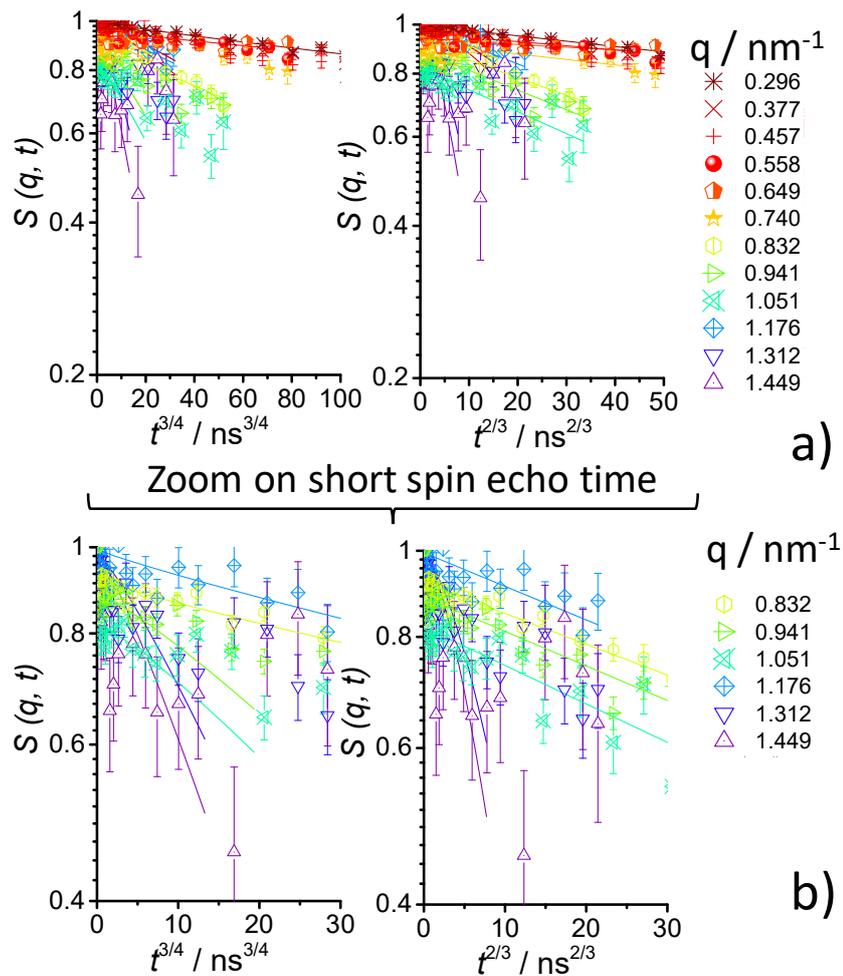

Figure S 6 – a) Plot of the spin-echo intermediate scattering function recorded for G-C18:1 Fiber gel as a function of stretched a) $t^{3/4}$ and a) $t^{2/3}$ exponentials of spin-echo time recorded at room temperature (25°C). b) A zoom at shorter spin echo time and q> 0.832 nm$^{-1}$ is given in b).